\documentclass[10pt,conference]{IEEEtran}

\pdfoutput=1
\usepackage{graphicx}
\usepackage{amsmath}
\usepackage{tabularx}  
\usepackage{multirow}  
\usepackage{booktabs}  
\usepackage{caption} 
\usepackage{url}
\usepackage{algorithm,algpseudocode}
\usepackage{amsfonts}  
\begin{document}

\title{SLIM: Saturation-Aware Lightweight Performance Modeling for LLM Serving}

\author{
\IEEEauthorblockN{Pol G. Recasens\IEEEauthorrefmark{1}\IEEEauthorrefmark{2}\IEEEauthorrefmark{3}, Ferran Agullo\IEEEauthorrefmark{1}\IEEEauthorrefmark{2}\IEEEauthorrefmark{3}, Yue Zhu\IEEEauthorrefmark{4}, \\ Chen Wang\IEEEauthorrefmark{4}, Jordi Torres\IEEEauthorrefmark{2}\IEEEauthorrefmark{3}, Josep Ll. Berral\IEEEauthorrefmark{3}\IEEEauthorrefmark{2}}
\IEEEauthorblockA{\IEEEauthorrefmark{2}Barcelona Supercomputing Center (BSC), \{pol.garcia, ferran.agullo, jordi.torres\}@bsc.es}
\IEEEauthorblockA{\IEEEauthorrefmark{3}Universitat Politècnica de Catalunya - BarcelonaTech (UPC), \{josep.ll.berral\}@upc.edu}
\IEEEauthorblockA{\IEEEauthorrefmark{4}IBM Research, \{Yue.Zhu, Chen.Wang1\}@us.ibm.com}}

\maketitle

\renewcommand{\thefootnote}{*}
\footnotetext{denotes equal contribution.}
\begin{abstract}
Large language model (LLM) serving commonly increases batch size to improve throughput, but performance eventually reaches a deployment-dependent plateau beyond which larger batches provide marginal gains while increasing latency and GPU memory consumption. Previous studies have attributed this behavior to HBM/DRAM bandwidth limitations, but the underlying causes have primarily been supported by conceptual arguments or high-level performance observations. As our first contribution, we present a detailed GPU characterization using hardware profiling techniques, demonstrating that throughput saturation originates in the attention kernels during the decode phase. Specifically, we show that their nearly constant arithmetic intensity as active-context lengths increases—not merely larger batch sizes—drives DRAM-bandwidth saturation, while the achieved compute throughput remains far below the hardware limit. Building on this analysis, we present the Batching Configuration Advisor (BCA), which selects the highest-throughput batching configuration satisfying a target latency constraint and identifies up to 55~GB of GPU memory allocation that can be avoided for the evaluated OPT models with minimal throughput loss. To enable these recommendations, we introduce SLIM (Saturation-Aware Lightweight Performance Model), a semi-analytical model that predicts LLM inference throughput and latency from analytical formulations of Transformer computation and memory traffic. Across the evaluated scenarios, SLIM outperforms representative performance-modeling baselines while successfully generalizing to previously unseen operating conditions.

\end{abstract}

\section{Introduction}

Large language models (LLMs) have become commoditized and are now widely used across personal and professional settings, supporting applications ranging from chatbots and voice interfaces to writing assistants and code-generation tools~\cite{bick2025rapid, humlum2024adoption, Liang2025}. Despite their widespread adoption, the computational requirements of state-of-the-art LLMs remain beyond the capabilities of most end users, making cloud-based inference the dominant deployment model. Consequently, service providers must efficiently accommodate large numbers of concurrent requests exhibiting diverse characteristics, including varying context lengths, arrival rates, and latency requirements~\cite{298685, jaiswal2025sageserve}. At the same time, these providers are transitioning to heterogeneous hardware architectures with task-specific accelerators, while serving models of varying complexity across user tiers, further increasing system heterogeneity~\cite{li2024large, jaiswal2025sageserve}.

A key challenge in online LLM serving is understanding how this heterogeneity translates into accelerator-level performance. Among the factors affecting serving efficiency, the incoming request rate is particularly important because it determines the degree of batching that can be exploited during inference. Increasing batch size generally improves throughput by exposing greater request-level parallelism, albeit at the cost of higher request latency. However, throughput eventually reaches a saturation point beyond which additional parallelism provides only marginal performance gains, while latency continues to increase. At the same time, GPU memory consumption grows because additional requests require larger aggregate KV caches. Although this phenomenon, commonly referred to as the \textit{throughput plateau}, has been widely observed in LLM serving systems, its underlying causes remain insufficiently explored.

Several studies attribute this throughput saturation under large-batch workloads to the increasing cost of KV cache memory transfers, which are identified as the primary performance bottleneck ~\cite{pope2023efficiently,davies2025efficient}. However, these conclusions are generally based on conceptual reasoning or on high-level performance observations, and lack detailed kernel-level analysis of the underlying execution behavior. In this work, we perform an in-depth characterization of LLM inference throughput using GPU kernel profiling and hardware-level analysis, demonstrating with low-level performance metrics that the throughput plateau is fundamentally caused by DRAM-bandwidth saturation in the attention kernels, resulting from excessive memory traffic during the decode phase. Furthermore, we show that this limitation is governed not only by batch size but also by the \textit{active context}, defined as the total number of tokens involved at each processing step. Our analysis shows that matrix multiplication kernels increasingly benefit from larger active contexts, achieving higher hardware utilization as workload size grows. In contrast, attention kernels exhibit nearly constant arithmetic intensity, causing their bandwidth demands to increase proportionally with active context. Consequently, the memory-bandwidth demand of attention operations eventually saturates the available DRAM bandwidth, creating a critical knee point beyond which throughput can no longer scale.

Identifying this memory-driven saturation mechanism enables a more principled approach to accelerator management. Once DRAM bandwidth is saturated, further increases in batch size provide negligible throughput benefits despite consuming more GPU memory and increasing latency. Avoiding operating beyond this point can free GPU memory capacity for other workloads, model instances~\cite{recasens2025mind}, or service tiers, improving overall system efficiency. We therefore introduce the \textit{Batching Configuration Advisor} (BCA), a lightweight advisory framework that determines the most efficient operating point for a given accelerator. BCA selects the highest-throughput configuration that satisfies a target latency constraint and maintains a user-defined minimum batching efficiency. This criterion avoids operating deep within the throughput plateau, where additional concurrency provides diminishing returns while increasing latency and memory consumption. 

To support BCA, we further propose the \textit{Saturation-Aware Lightweight Performance Model} (SLIM), a semi-analytical performance model for LLM serving. SLIM leverages the performance insights obtained from our characterization study to estimate throughput and latency across the heterogeneous operating conditions typical of service providers. The model combines interpretable analytical formulations of the computation and memory traffic generated by Transformer operations with a small set of deployment-specific constants calibrated through a lightweight profiling phase. This lightweight calibration enables BCA to reduce profiling time by 43.85\% compared to exhaustive profiling. We compare SLIM against two state-of-the-art performance models, LLMVisor \cite{jinllmvisor} and the model proposed by Imai et al.~\cite{imai2024predicting}. SLIM consistently outperforms both models when generalizing to previously unseen operating conditions, including different LLMs and active-context lengths, reducing the mean absolute percentage error (MAPE) by an average of 79.3\% across all evaluated scenarios. Beyond its role in BCA, SLIM may provide broader value for online service providers by enabling informed capacity-planning decisions and more efficient hardware utilization. Its latency predictions support service-tier configuration and quality-of-service management, while throughput estimates facilitate resource provisioning by determining the accelerator capacity required to meet workload demands and performance targets.

In summary, our key contributions are:

\begin{itemize}

    \item \textbf{We demonstrate that DRAM-bandwidth saturation in attention kernels is the primary driver of throughput saturation under large active-context workloads}. Through detailed GPU-level characterization, we show that increasing the active context proportionally increases KV-cache memory traffic in the attention kernels, resulting in nearly constant arithmetic intensity. As a result, attention approaches the GPU memory-bandwidth limit, memory-related stalls increase, and compute resources remain underutilized, ultimately producing the throughput plateau.

    \item \textbf{We propose SLIM, a saturation-aware lightweight performance model for LLM serving.}
    \textsc{SLIM} combines a semi-analytical formulation with a lightweight profiling phase to accurately predict serving throughput and latency. It successfully generalizes to unseen operating conditions, including larger models and longer output lengths, achieving a throughput MAPE of 17.6\% MAPE under combined model-and-output-length generalization and substantially outperforming the evaluated baselines.
    
    \item \textbf{We introduce the Batching Configuration Advisor (BCA), a lightweight advisory framework for accelerator operation.} BCA selects the highest-throughput configuration that satisfies a target latency constraint and maintains a user-defined minimum batching efficiency. Experimental results show that BCA identifies configurations that reduce memory requirements by up to 55 GB for the tested OPT models, while requiring 43.85\% less profiling time than exhaustive configuration search.
    
\end{itemize}

\textit{Paper structure}: The remainder of this paper is organized as follows. Sections~\ref{sec:background} and~\ref{sec:relatedwork} provide the necessary background and review the related work, with the former also illustrating the throughput plateau in practice. Section~\ref{sec:methodology} describes the methodology and experimental setup used throughout the paper. The three main contributions of this work are then presented and evaluated in separate sections. Section~\ref{sec:gpuprofiling} presents the low-level performance analysis that identifies the causes of throughput saturation under large active-context workloads. Section~\ref{sec:modeling} introduces \textsc{SLIM}, the proposed performance model, and Section~\ref{sec:advisor} presents the configuration advisor for optimizing LLM serving. Finally, Sections~\ref{sec:discussion} and~\ref{sec:conclusion} discuss the implications of the results and conclude the paper.

\section{Background}\label{sec:background}

\subsection{LLM Inference and KV Caching}
Modern large language models such as OPT~\cite{zhang2022OPT}, GPT~\cite{brown2020language}, and Llama~\cite{touvron2023Llama} are trained using the next-token prediction objective, where output tokens are generated autoregressively conditioned on a prompt \(x\). During inference, request processing consists of two phases. First, in the \textit{prefill} phase, all prompt tokens are processed in parallel to generate the first output token. Subsequently, the model enters the \textit{decode} phase, where new tokens are generated sequentially by conditioning on previously generated tokens. Generation terminates when an end-of-sequence token is produced or a predefined maximum length is reached.

These models are built upon the Transformer architecture~\cite{vaswani2017attention}, which stacks multiple layers composed of a self-attention mechanism and a feed-forward network (FFN). Given an input representation \(X\), the attention module first projects the input into queries \(Q\), keys \(K\), and values \(V\) using learnable weight matrices \(W_Q\), \(W_K\), and \(W_V\):
\begin{equation}
Q = XW_Q,\qquad
K = XW_K,\qquad
V = XW_V.
\end{equation}

It then computes the output as:
\begin{equation}
\mathrm{Attention}(Q,K,V)
=
\mathrm{softmax}
\left(
\frac{QK^\top}{\sqrt{d_k}}
\right)V
\end{equation}

The resulting representations are then processed by the FFN, with residual connections and normalization applied around these sublayers.

To avoid recomputing attention states for previously processed tokens during autoregressive decoding, modern LLM serving systems maintain a key-value (KV) cache in GPU memory. The KV cache stores the key and value tensors generated by each Transformer layer and allows attention computation for a newly generated token to reuse previously computed states. Consequently, the attention operation during decoding is reduced from a matrix-matrix multiplication \(QK^T\) over the entire sequence to a matrix-vector multiplication \(qK^T\), where \(q\) is the query corresponding to the current token.

\subsection{Memory Bottlenecks and Batching}
Although the KV cache significantly reduces the computational cost of autoregressive generation, it introduces a memory-access bottleneck during the decode phase. For each newly generated token, the key and value tensors associated with all previously processed tokens must be retrieved from GPU main memory (HBM/DRAM) and transferred to the on-chip memory hierarchy used by GPU execution kernels. Consequently, decoding exhibits low \emph{arithmetic intensity}, defined as the ratio of floating-point operations (FLOPs) to bytes transferred from memory. Workloads with low arithmetic intensity are memory-bound, meaning that performance is constrained primarily by memory bandwidth rather than computational throughput.

To alleviate the memory-bound nature of decoding, modern serving systems process multiple requests concurrently through batching~\cite{recasens2025mind}, increasing the amount of computation performed per decoding step. However, traditional static batching is inefficient for LLM workloads, as requests can generate outputs of different lengths, causing some batch slots to remain idle while others continue decoding. To address this limitation, state-of-the-art serving systems employ \emph{continuous batching}~\cite{yu2022orca, kwon2023efficient, tensorrt-llm, deepspeed-mii}, which dynamically admits new requests and removes completed ones between decoding iterations, maintaining high batch occupancy.

The maximum batch size is ultimately constrained by GPU memory capacity. Since each active request maintains a growing KV cache, memory consumption increases with both sequence length and concurrency. To improve memory efficiency, modern serving frameworks such as vLLM~\cite{kwon2023efficient} and S-LoRA~\cite{sheng2024slora} allocate KV-cache memory incrementally instead of reserving space for the maximum output length upfront, enabling higher request concurrency and better resource utilization.

\subsection{Throughput Saturation and the Latency Trade-off}
\begin{figure}[ht]
\centering
  \vspace{-5pt}
  \includegraphics[width=0.8\linewidth]{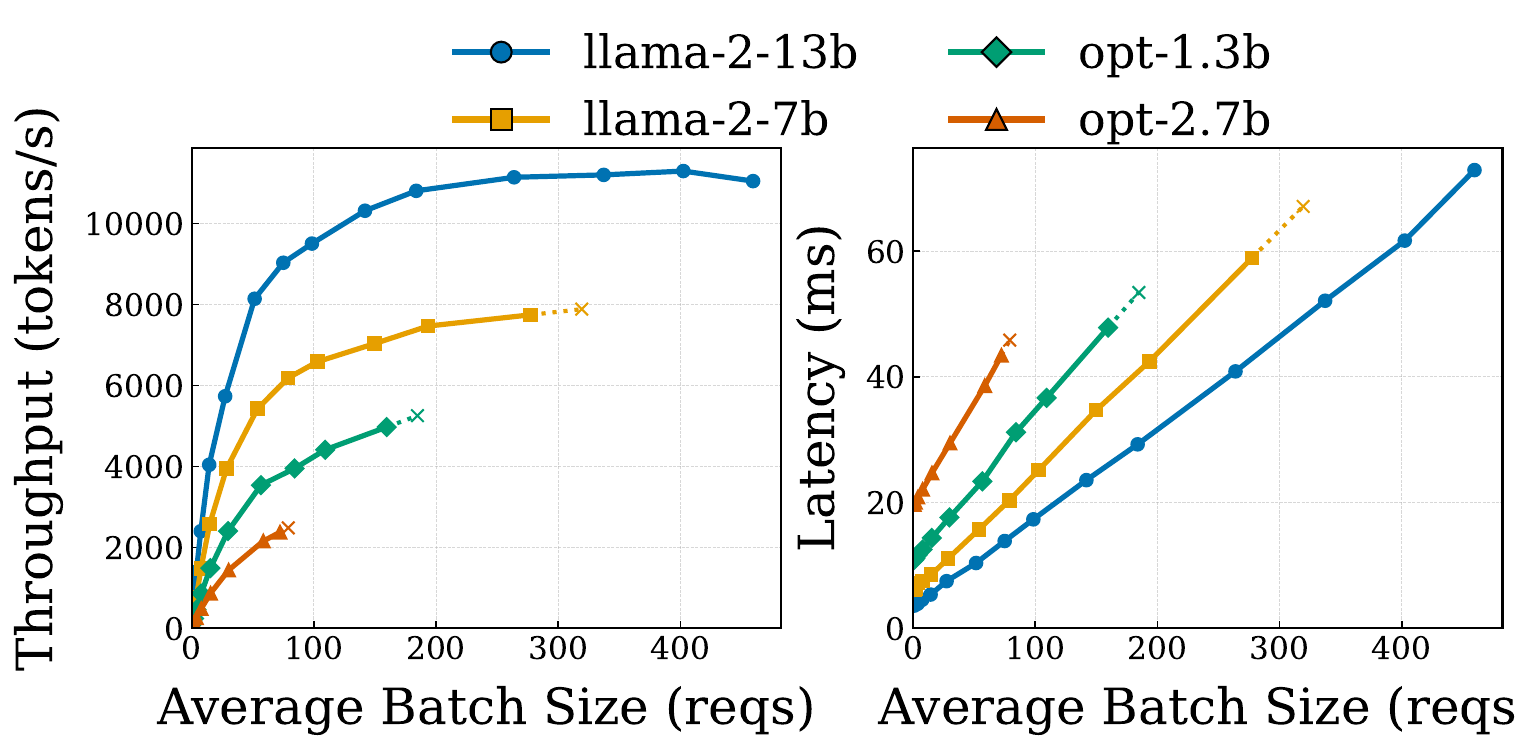}
  \caption{Throughput (tokens/s) and latency (ms) evolution when setting the maximum batch size to values in range 1..512 across different models (OPT-1.3B, OPT-2.7B, Llama-2-7B and Llama-2-13B). The X-axis corresponds to the average batch size, rather than the configured maximum, and the crosses mark the point at which the KV cache capacity is exceeded due to the increased batch size. Results are obtained in the online mode described in Section~\ref{sec:methodology}.}
  \label{fig:background_throughput_latency}
\end{figure}

As discussed above, batching improves the performance of LLM serving systems by allowing multiple requests to be processed simultaneously. As illustrated on the left side of Figure~\ref{fig:background_throughput_latency}, throughput increases with the batch size across all evaluated models. These gains, however, come at the cost of higher latency. Processing more requests in parallel increases overall system efficiency and throughput, but also prolongs the execution time experienced by each request. Depending on the target application, this throughput--latency trade-off may be acceptable or even desirable.

However, throughput improvements diminish well before the GPU memory capacity is exhausted (marked by crosses in the figure), a phenomenon commonly referred to as the \emph{throughput plateau}. For small batch sizes, throughput scales almost linearly with the number of requests processed concurrently. Beyond a certain point, however, the rate of improvement decreases significantly and eventually saturates, while latency continues to grow approximately linearly. As a result, allocating additional memory to support larger batches yields limited throughput benefits while incurring increasingly higher response times and GPU memory usage.

\section{Related Work}\label{sec:relatedwork}

\subsection{LLM Inference Performance Analysis} 
The performance of autoregressive LLM inference at low batch sizes is widely acknowledged to be constrained by memory bandwidth rather than computational throughput. While the prefilling phase is executed only once per inference request, the decoding phase, characterized by low arithmetic intensity, is repeatedly invoked for each generated token, making memory accesses the dominant factor affecting inference performance~\cite{yuan2024llm,kwon2023efficient}. This bottleneck is evidenced by the substantial body of work dedicated to increasing the arithmetic intensity of decoding. Representative approaches include chunked prefilling and speculative decoding, which improve hardware utilization by increasing the amount of computation performed per memory access~\cite{agrawal2024taming,leviathan2023fast,chen2023accelerating}. Other techniques, such as multi-query and grouped-query attention~\cite{shazeer2019fast,ainslie2023gqa}, as well as quantization methods~\cite{lin2024awq,frantar2022gptq}, seek to reduce KV cache memory traffic, thereby alleviating memory bandwidth constraints during inference.

Nevertheless, limited attention has been devoted to understanding the performance limits of LLM inference at large batch sizes and active context lengths. In particular, the fundamental causes of the throughput plateau regime discussed previously remain insufficiently characterized. Several studies suggest that KV-cache memory transfers become the primary bottleneck at large batch sizes ~\cite{pope2023efficiently,davies2025efficient}. However, these explanations remain conceptual without a detailed low-level characterization. Consequently, the interplay between arithmetic intensity, memory traffic, and throughput saturation remains poorly understood and has not been systematically characterized at the kernel level. In this work, we take a step toward addressing this gap by providing a low-level analysis of modern LLM inference workloads under large batch sizes and active context lengths. Leveraging kernel-level measurements and roofline-based visualizations, we systematically analyze the factors governing throughput saturation and reveal the underlying causes of the observed plateau regime.

\subsection{Performance Modeling} 

Recent studies have proposed performance models and simulators for LLM serving, although they target different optimization objectives. Vidur~\cite{agrawal2024vidur} is a large-scale simulation framework that models LLM inference operators using a combination of experimental profiling and predictive modeling, enabling end-to-end estimation of serving metrics across different workloads and deployment configurations. GenZ~\cite{bambhaniya2024demystifying} models the compute, memory, and communication requirements of LLM inference across model, workload, and platform configurations, while LLMCompass~\cite{zhang2024llmcompass} combines operator-level performance modeling with mapping exploration to evaluate accelerator designs. Similarly, AMALI~\cite{cao2025amali} analytically models LLM inference on modern GPUs by accounting for architectural execution characteristics. 

Imai et al.~\cite{imai2024predicting} introduce a latency prediction model that combines roofline-based analytical estimates with regression models trained on serving information, but focus on small batch sizes and input lengths, without evaluating the model in the throughput saturation regime. Similarly, LLMVisor~\cite{jinllmvisor} introduces a roofline-guided latency attribution model for multi-tenant LLM serving, decomposing batch latency into per-request costs based on computation and memory-access features. These approaches are useful for simulating, estimating, or attributing serving performance, but they do not explicitly model the onset of the throughput plateau caused by decode-time attention bandwidth saturation or translate it into an active-context capacity recommendation. Sheng et al.~\cite{sheng2024fairness} propose Virtual Token Counter (VTC), a fair scheduling policy that approximates service through token-based costs. However, token-count abstractions do not capture hardware bandwidth limits or the dependence of attention traffic on batch size and context length. Unlike these, SLIM extrapolates to unseen model scales and context lengths from a fixed calibration, without refitting. This enables SLIM to capture the hardware conditions under which batching stops providing proportional throughput gains, enabling knee-point estimation without exhaustively profiling the complete throughput curve.

\subsection{Conference Extension}
A preliminary version of this work was presented at the IEEE International Conference on Cloud Computing~\cite{recasens2025mind}. This journal article substantially extends that prior work in two key directions. First, it generalizes the low-level analysis of throughput saturation from the batch-size dimension to large active-context scenarios, providing a more comprehensive characterization of the memory and computational bottlenecks that arise in modern LLM-serving workloads. Additionally, all evaluations have been repeated using a newer version of vLLM and a more recent set of LLMs. Second, it introduces SLIM, a performance model capable of predicting serving throughput and latency while requiring significantly fewer profiling measurements and generalizing effectively to previously unseen operating conditions. By enabling efficient exploration of the configuration space without exhaustive benchmarking, SLIM substantially reduces the profiling overhead associated with system tuning and serves as the foundation of the proposed BCA advisor.

\section{Methodology}
\label{sec:methodology}

We conduct our experiments using the vLLM framework~\cite{kwon2023efficient}, version 0.15.1. We use vLLM in two modes. In \textbf{online mode}, we follow a client-server architecture and send HTTP requests to a local server instance. In \textbf{offline mode}, used for the low-level GPU profiling in Section~\ref{sec:gpuprofiling} with Nsight Systems and Nsight Compute, we instantiate vLLM directly in Python, injects synthetic requests into the scheduler and then runs the prefill and decode phases with explicit \textit{llm\_engine.step()} calls. We retain the vLLM default setting but disable vLLM log statistics and prefix caching for controlled profiling runs. 

\textbf{Workload.} For the performance analysis, we generate synthetic prompts and independently vary the batch size, input length, and output length to isolate the impact of each factor on system performance. For the remaining evaluations, scenarios requiring fixed input and output lengths are generated using the random dataset generator provided by vLLM. In contrast, evaluations targeting realistic serving conditions use requests sampled from the cleaned ShareGPT dataset~\cite{ShareGPT_Vicuna_unfiltered}, thereby preserving the heterogeneous input and output length distributions observed in real-world workloads.

\textbf{Models.} For the performance analysis, we evaluate two representative LLMs: Mistral-7B~\cite{jiang2023mistral7b} and Granite-8B~\cite{mishra2024granite}. For all remaining experiments, we employ models from the OPT family~\cite{zhang2022optopenpretrainedtransformer}, including OPT-125M, OPT-350M, OPT-1.3B, OPT-2.7B, and OPT-6.7B. In addition, for large-scale multi-GPU settings, we employ models from the Qwen family~\cite{qwen2}, specifically Qwen-32B and Qwen-72B.

\textbf{Hardware.} All experiments run on nodes equipped with four NVIDIA Hopper H100 (64GB HBM2), 512GB RAM memory, and 80 CPU cores. Single-GPU experiments request one GPU, while the Qwen-32B and Qwen-72B online experiments request two and four GPUs and use tensor parallelism. The launcher allocates 20 CPU cores per requested GPU. Accordingly, the SLIM performance model is parameterized using the configured peak hardware capabilities of the H100 GPU, namely 989 TFLOP/s of tensor-core computational throughput and 1.62 TB/s of memory bandwidth.

\section{Performance Analysis}
\label{sec:gpuprofiling}

In this section, we present a low-level analysis of the throughput plateau observed under large active-context workloads. Unlike prior studies, our work is the first to provide a comprehensive explanation of this performance bottleneck using detailed GPU tracing data. We first examine how the different workload dimensions that determine the active context contribute to throughput saturation. We then investigate its underlying causes through a hierarchical analysis, beginning with the execution-time contributions of the prefill and decode phases and concluding with a kernel-level characterization of attention execution.

Our findings reveal that DRAM-bandwidth saturation in attention kernels is the primary cause of the throughput plateau in large active-context scenarios. We observe that their arithmetic intensity remains nearly constant across increasing input lengths, output lengths, and batch sizes. This ultimately leads to memory-bandwidth saturation, leaving a significant portion of computational resources underutilized.

\subsection{Impact of Active Context in Serving Performance}

We begin by revisiting the throughput plateau introduced in Figure~\ref{fig:background_throughput_latency}. Whereas prior analyses have primarily examined this phenomenon as a function of batch size, we extend the characterization to include both input and output sequence lengths. Together, these dimensions determine the \emph{active context} $A_t$, defined as the total number of tokens processed by the model at a given inference step. Formally, for a batch $\mathcal{B}_t$ at processing step $t$, we define the active context as:
\begin{equation}
A_t = \sum_{r \in \mathcal{B}_t} \left(IL_r + y_{r,t}\right)
\end{equation}
where $IL_r$ is the input length of request $r$ and $y_{r,t}$ is the number of tokens already generated for that request.

\begin{figure}[ht]
  \centering
  \includegraphics[width=1\linewidth]{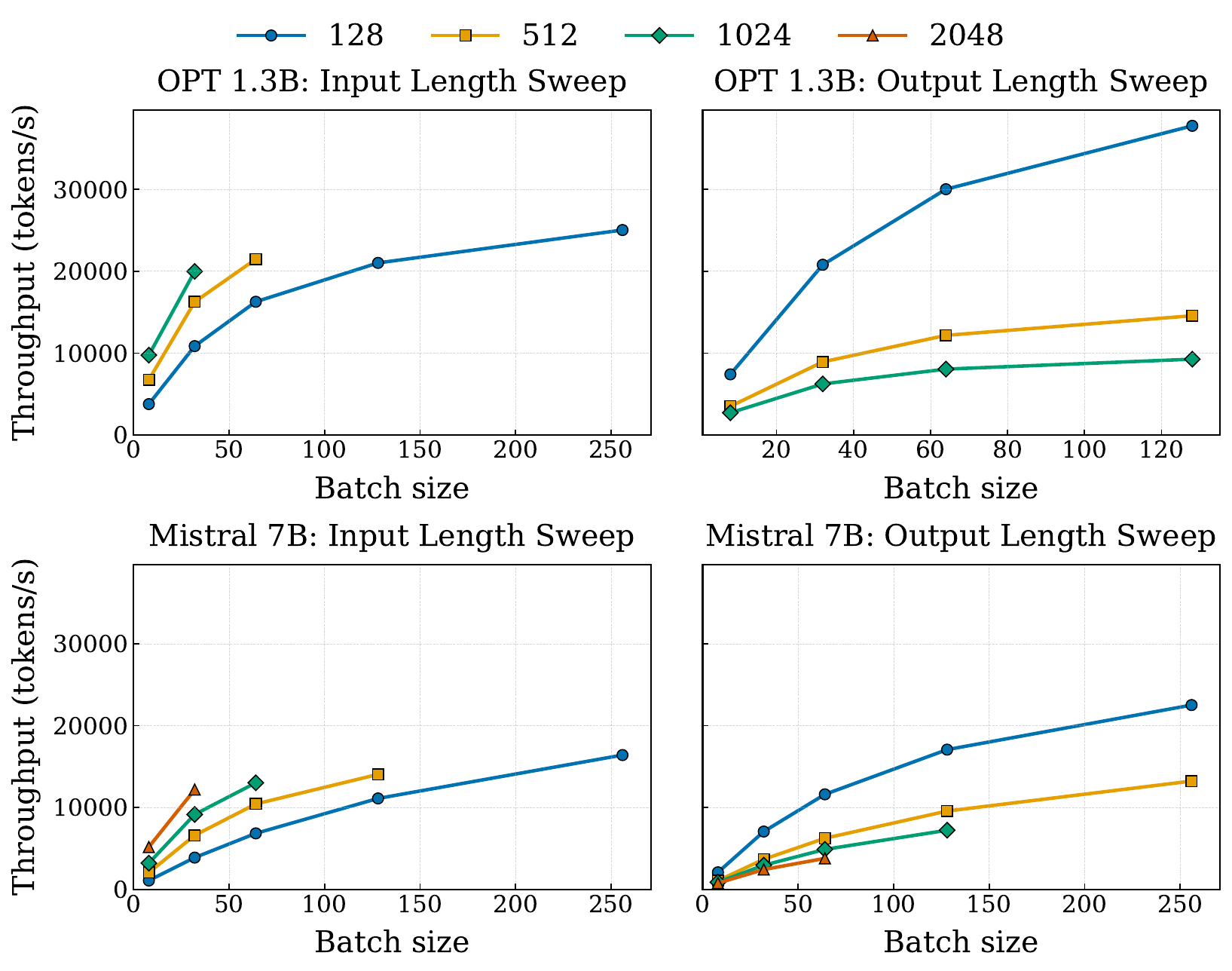}
  \caption{Throughput evolution under varying input lengths (left) and output lengths (right) across different batch sizes. The top row corresponds to OPT-1.3B, while the bottom row presents the results for Mistral-7B. Missing data points at larger batch sizes correspond to out-of-memory failures, as the available GPU memory was insufficient to accommodate the increased KV-cache footprint associated with longer active contexts.}
  \label{fig:throughput_latency_context}
\end{figure}

Figure~\ref{fig:throughput_latency_context} shows how throughput saturation evolves with batch size, input length, and output length. Although all configurations eventually reach a plateau, their saturation profiles differ, indicating that the plateau is governed not only by batch size but also by the total active context processed by the model. Increasing input length generally improves throughput because a larger fraction of execution is spent in the highly parallel prefill phase. In contrast, increasing output length extends the less efficient autoregressive decode phase, in which each generated token attends to a progressively longer context, thereby reducing execution efficiency and lowering throughput.

\subsection{System-level Analysis} \label{sec:decode_prefill}

\begin{figure}[ht]
  \centering
  \includegraphics[width=1.0\linewidth]{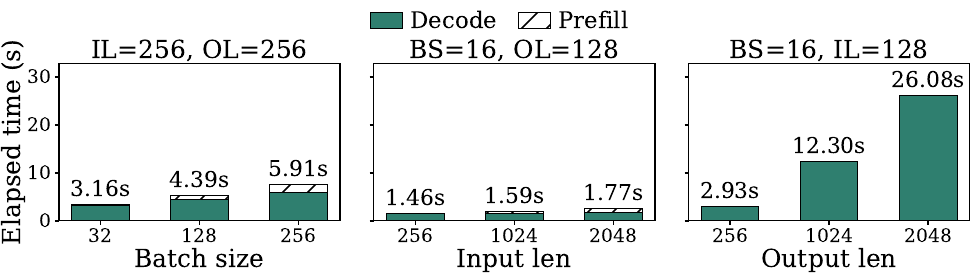}
  \caption{Breakdown of total inference time into prefill and decode phases for the \textit{Mistral-7B} model under increasing batch size, input length, and output length.}
  \label{fig:prefill_decode}
\end{figure}

We continue the analysis by decomposing average inference time into the prefill and decode phases and examining how their contributions vary with batch size, input length, and output length. As shown in Figure~\ref{fig:prefill_decode}, decode consistently dominates total execution time across all evaluated scenarios, making it the primary contributor to inference latency and the most likely source of the observed throughput plateau.

Although the relative contribution of prefill increases with batch size and input length, it remains substantially smaller than that of decode, even for the largest configurations. Increasing input length raises total execution time modestly (below 2s), because the additional prefill computation is highly parallelizable. In contrast, increasing the output length has a much larger effect, extending the execution time to 26s for 2048 generated tokens. As output length increases, the prefill contribution becomes negligible, further confirming decode as the dominant component of end-to-end inference time.

\begin{figure}[ht!]
  \centering
  \includegraphics[width=1.0\linewidth]{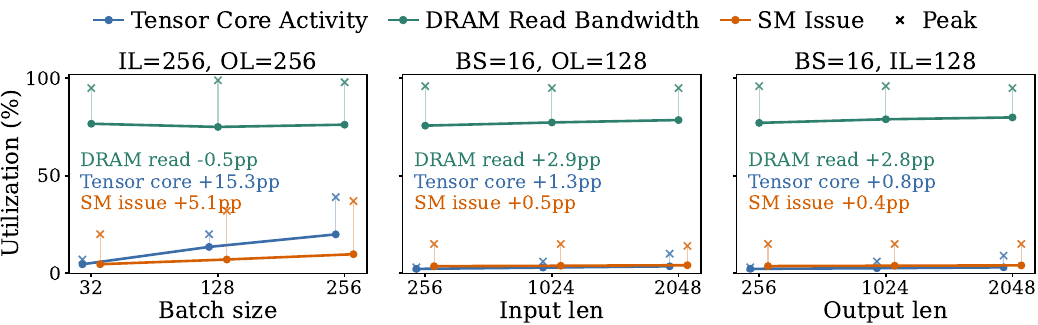}
\caption{Evolution of average Tensor Core activity, SM instruction-issue throughput, and DRAM read-bandwidth utilization during the decode phase of Mistral-7B as batch size, input length, and output length increase. Crosses denote the corresponding peak values observed during each execution.}
  \label{fig:decode_kernels_batch_size_evolution}
\end{figure}

In light of these observations, we focus the remainder of our analysis on the decode phase. Figure~\ref{fig:decode_kernels_batch_size_evolution} presents GPU execution metrics for the full decode phase across workloads with varying batch sizes, input lengths, and output lengths. We consider three complementary metrics. First, \emph{Tensor Core Activity (\%)} measures the utilization of the Tensor Core execution units, capturing the activity of the dense matrix multiplication operations that dominate Transformer execution. Second, \emph{SM Issue (\%)} quantifies the achieved instruction-issue throughput across the SM warp schedulers, providing a broader view of execution activity beyond operations executed on the Tensor Cores. Finally, \emph{DRAM Read Throughput} measures the fraction of available memory bandwidth used to transfer data from GPU main memory to the on-chip cache hierarchy, providing a direct indication of memory pressure during execution.

Overall, GPU execution remains far from its Tensor Core and instruction-issue limits, with average Tensor Core activity below 30\% and average SM Issue below 10\% across all evaluated configurations. In contrast, average DRAM read-bandwidth utilization remains close to 80\% and frequently reaches peaks near 100\%, providing strong evidence that throughput saturation under large active-context workloads is primarily driven by memory-bandwidth limitations. At batch size 256, which already lies within the throughput plateau as shown in Figure~\ref{fig:throughput_latency_context}, DRAM bandwidth is near saturation while Tensor Core activity and SM Issue remain substantially below their peak levels. Across the three scenarios, DRAM-bandwidth utilization changes by only $-0.5$, 2.9, and 2.8 percentage points because it is already close to its limit. These results indicate that increasing the active context primarily intensifies memory pressure without producing a comparable increase in Tensor Core activity or instruction-issue throughput.

\subsection{Kernel-level Analysis} \label{sec:attention_kernel}

\begin{figure}[ht]
  \centering
  \includegraphics[width=1\linewidth]{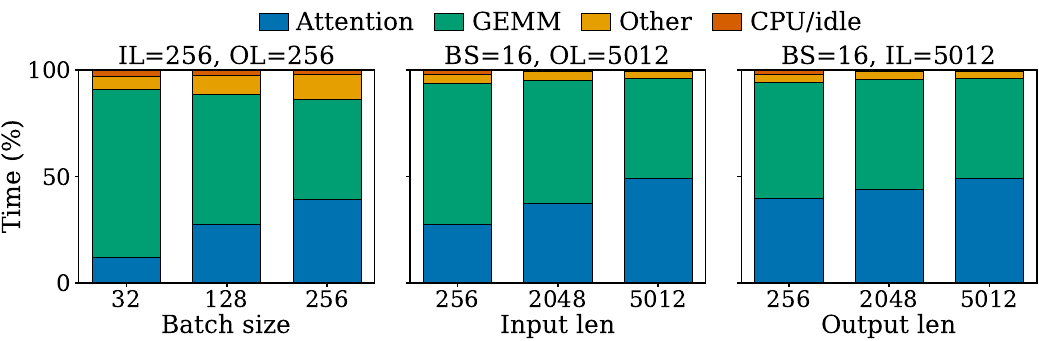}
  \caption{Contribution of GPU kernels to the execution time of the decode phase as the active context increases for Mistral-7B. The execution times of the attention and GEMM kernels (corresponding to the FFN block) are reported separately as the dominant contributors, while the remaining kernels are aggregated under \textit{Other}.}
  \label{fig:decode_kernels_distinct_kernels}
\end{figure}

To further investigate the decode phase, we analyze how its execution time is distributed across GPU kernels. Figure~\ref{fig:decode_kernels_distinct_kernels} reports the average contribution of each kernel category. Decode time is dominated by the attention mechanism and the GEMM operations associated with the feed-forward network (FFN), the two main computational components of the Transformer block. As the active context increases, attention accounts for a progressively larger share of decode time, regardless of whether this increase is driven by batch size, input length, or output length. Moreover, in the batch-size regime where the throughput plateau emerges (i.e., batch sizes greater than 64, as shown in Figure~\ref{fig:throughput_latency_context}), attention becomes the dominant execution component, exceeding the contribution of FFN computations. These results identify attention as the primary source of the throughput plateau under large active-context workloads.

\begin{figure}[ht]
  \centering
  \vspace{-5pt}
  \includegraphics[width=1.0\linewidth]{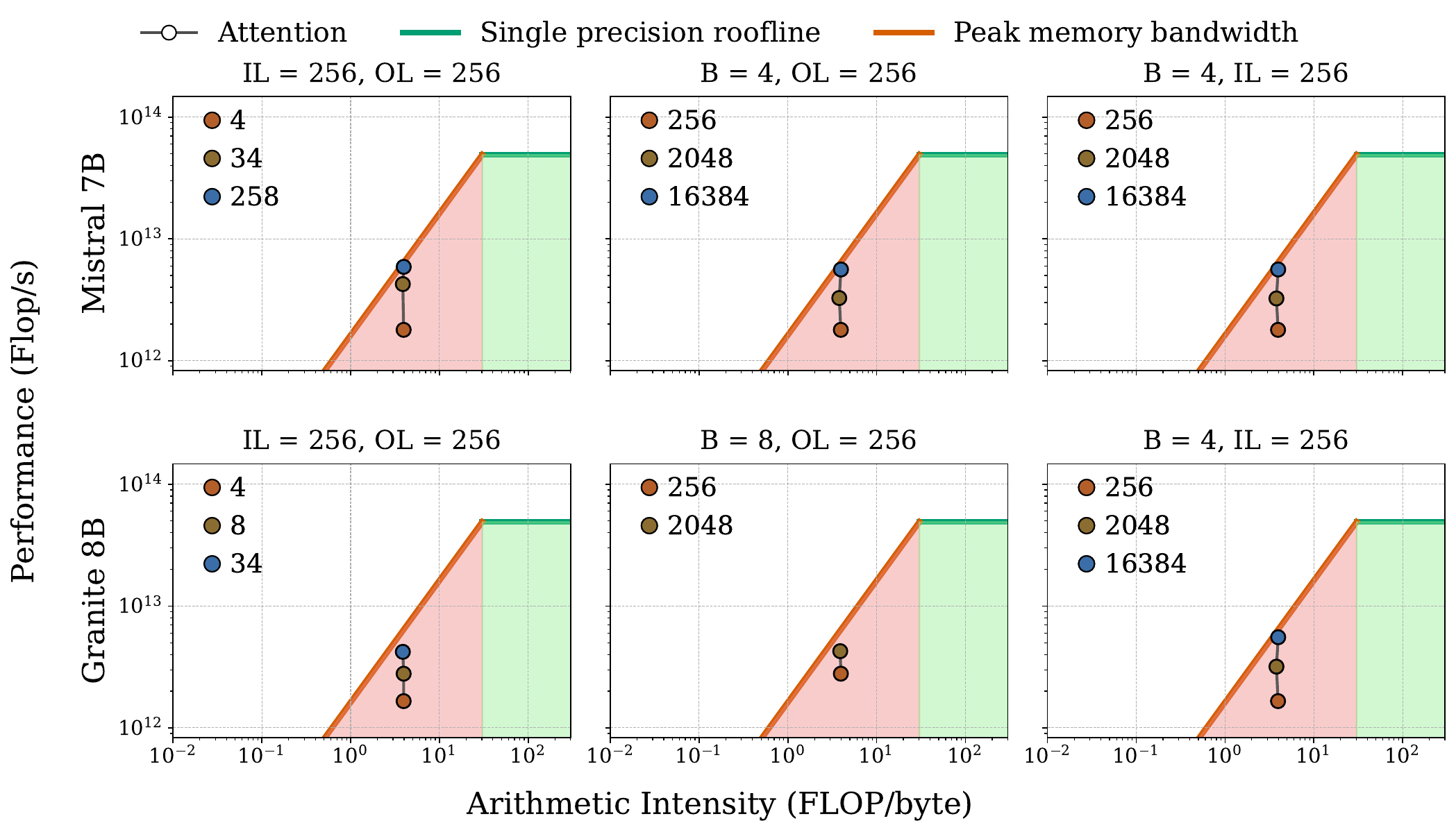}
  \caption{Roofline characterization of the attention kernels executed during the final decode step under increasing batch size, input length, and output length. The top row corresponds to Mistral-7B, while the bottom row presents the results for Granite-8B.}
  \label{fig:roofline_comparison}
\end{figure}

We next examine the attention kernel in greater detail. Figure~\ref{fig:roofline_comparison} presents its roofline characterization across increasing active-context lengths for Mistral-7B and Granite-8B. The roofline model relates kernel performance to arithmetic intensity (AI), defined as the number of arithmetic operations performed per byte transferred from DRAM. Kernels with low AI operate in the memory-bound regime (red), where performance is limited by memory bandwidth, whereas kernels with high AI operate in the compute-bound regime (green), where performance is limited by computational throughput. The roofline boundaries are defined by the device's peak single-precision compute throughput and peak DRAM bandwidth, representing the accelerator's theoretical maximum compute performance and memory-transfer capability, respectively.

The complete quantitative results are reported in Table~\ref{tab:decode-attention-roofline-evidence}. As shown in Figure~\ref{fig:roofline_comparison}, the attention kernel remains in the memory-bound regime across all evaluated configurations, with arithmetic intensity varying only marginally as the active context increases. Furthermore, GPU compute capacity also remains largely underutilized. Across all evaluated configurations, the kernel sustains less than 6 TFLOP/s, more than two orders of magnitude below the hardware’s theoretical peak Tensor Core throughput of 989~TFLOP/s. These results demonstrate that increasing the active-context size \textemdash whether by increasing batch size, input length, or output length\textemdash does not shift the attention kernel toward the compute-bound regime, because the additional computation is accompanied by a proportional increase in DRAM traffic.

Accordingly, as the active context increases, the attention kernel follows an almost vertical trajectory in the roofline model, reflecting its nearly constant arithmetic intensity while performance approaches the DRAM-bandwidth ceiling. Memory bandwidth therefore becomes the limiting resource, preventing further performance gains once saturation is reached. This behavior is evident at batch size 256, which lies within the throughput plateau regime: the attention kernel operates near the roofline's memory-bandwidth limit and sustains approximately 90\% average DRAM-bandwidth utilization. Although this average is computed over the full kernel execution, instantaneous utilization reaches 100\%, as shown in Figure~\ref{fig:decode_kernels_batch_size_evolution}. The same behavior also emerges as input and output lengths increase, since all three workload dimensions raise memory traffic while leaving attention arithmetic intensity essentially unchanged. These results identify DRAM-bandwidth saturation as the primary cause of the attention-kernel performance bottleneck, explaining the increasing slowdown observed in Figure~\ref{fig:decode_kernels_distinct_kernels} under large active contexts.

\begin{table*}[t]
\footnotesize
\centering
\caption{Quantitative results corresponding to Figure~\ref{fig:roofline_comparison} for Mistral-7B, reporting roofline measurements of the attention kernel during the final decode step under increasing batch size, input length, and output length. The active context is defined as $B(IL+OL)$, where $B$ is the batch size, $IL$ the input length, and $OL$ the output length.}
\label{tab:decode-attention-roofline-evidence}
\begin{tabular}{llrrrr}
\toprule
Sweep & Configuration & Active context & Arithmetic Intensity & Performance & DRAM Avg. Utilization \\
 & $(B,IL,OL)$ & (tokens) & (FLOP/byte) & (TFLOP/s) & (\%) \\
\midrule
Batch size & $(4,256,256)$ & 2,048 & 3.95 & 1.79 & 28.0 \\
Batch size & $(34,256,256)$ & 17,408 & 3.87 & 4.25 & 67.9 \\
Batch size & $(258,256,256)$ & 132,096 & 3.96 & 5.89 & 91.9 \\
\addlinespace
Input length & $(4,2048,256)$ & 9,216 & 3.79 & 3.27 & 53.2 \\
Input length & $(4,16384,256)$ & 66,560 & 3.97 & 5.61 & 87.2 \\
\addlinespace
Output length & $(4,256,2048)$ & 9,216 & 3.79 & 3.24 & 52.8 \\
Output length & $(4,256,16384)$ & 66,560 & 3.97 & 5.61 & 87.3 \\
\bottomrule
\end{tabular}%
\end{table*}

\subsection{Low-level Memory Transfer Analysis}
We conclude the analysis by examining two additional aspects of the attention kernel's memory behavior. Specifically, we analyze access locality across the GPU cache hierarchy and the fraction of memory-related stalls. Together, these metrics provide further evidence of DRAM-bandwidth saturation.

\begin{figure}[ht]
  \centering
  \vspace{-5pt}
  \includegraphics[width=1.0\linewidth]{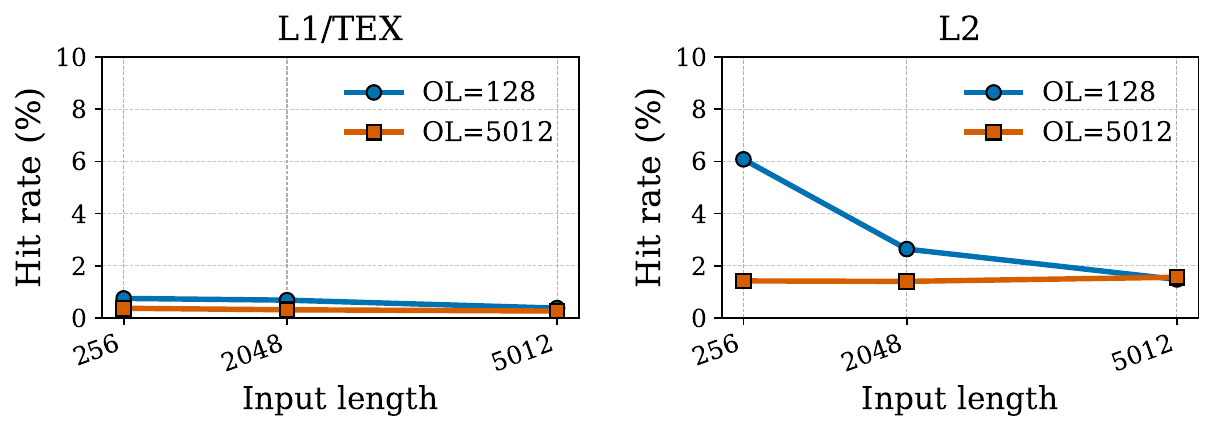}
  \caption{L1/TEX and L2 cache hit rates of the attention kernels executed during the final decode step of Mistral-7B as input length increases at a fixed batch size of 8. Each curve corresponds to a different output length.}
  \label{fig:cache_misses}
\end{figure}

Figure~\ref{fig:cache_misses} reports the hit rates of the L1/TEX and L2 caches. These caches exploit temporal and spatial locality by serving memory requests before they reach DRAM. Since data is transferred in cache lines rather than individual bytes, effective locality enables data reuse and reduces DRAM traffic. However, as shown in Figure~\ref{fig:cache_misses}, both cache levels exhibit consistently low hit rates across the evaluated active-context sizes. The L1/TEX hit rate remains below 1\%, while the L2 hit rate never exceeds 7\%. Moreover, cache locality also decreases as the active context grows. These results indicate limited data reuse in the attention kernel, which further deteriorates as the context length increases, leading to repeated DRAM accesses and greater memory-bandwidth pressure, consistent with previous observations.

\begin{figure}[ht]
  \centering
  \vspace{-5pt}
  \includegraphics[width=1.0\linewidth]{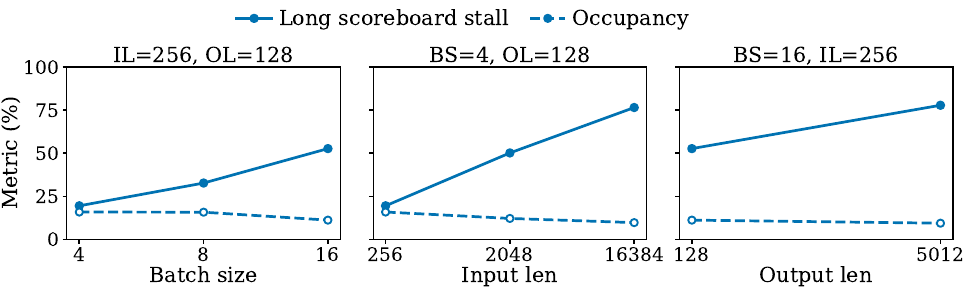}
    \caption{Percentage of \textit{Occupancy} and \textit{Long Scoreboard Stalls} for the attention kernels executed by Mistral-7B during the final decode step under increasing batch size, input length, and output length.} 
    \label{fig:attention_stalls}
\end{figure}

Finally, Figure~\ref{fig:attention_stalls} reports the \textit{Long Scoreboard Stalls}, which represent the percentage of active warps stalled per cycle on memory operations from the GPU memory hierarchy (L1/TEX, L2, or DRAM), and the \textit{Occupancy}, which reports the percentage of active warps resident on a Streaming Multiprocessor (SM) relative to the hardware maximum. As the active context increases, we observe that the proportion of long scoreboard stalls increases substantially, ultimately accounting for more than 50\% of active warp cycles in the largest active-context configurations. On the other hand, occupancy remains constant and below 20\%, suggesting limited latency-hiding capacity. Together with the previously observed DRAM-bandwidth saturation and low Tensor Core and SM utilization, these results indicate that many resident warps remain blocked on memory accesses while computational resources are underutilized.

\subsection{Concluding Remarks}

Collectively, these results show that the throughput plateau under large active-context workloads is driven by DRAM-bandwidth saturation in attention kernels. As the active context increases, their arithmetic intensity remains nearly constant, causing memory traffic to scale with computation until the available DRAM bandwidth is saturated, as demonstrated by the roofline analysis in Figure~\ref{fig:roofline_comparison}. Because attention dominates decode time at large active contexts (Figure~\ref{fig:decode_kernels_distinct_kernels}), and decode itself dominates overall execution time in this regime (Figure~\ref{fig:prefill_decode}), this bottleneck propagates to the end-to-end throughput and produces the observed throughput plateau. The low cache hit rates in Figure~\ref{fig:cache_misses} and the increasing memory-related stalls in Figure ~\ref{fig:attention_stalls} provide further evidence of this behavior.

\section{Performance Modeling}
\label{sec:modeling}

In this section, we present SLIM, a semi-analytical model for predicting LLM inference throughput and latency. SLIM builds on analytical formulations of the computation and memory traffic incurred by Transformer operations during the prefill and decode phases. These formulations are simplified to improve interpretability while preserving predictive accuracy, and are complemented with a small set of constants calibrated from empirical profiling data.

A key advantage of SLIM is that it achieves accurate predictions from a small set of profiling measurements while generalizing to unseen execution conditions. This capability stems from explicitly modeling the key factors governing inference performance, including model architecture, sequence length, and batch size, rather than relying solely on data-driven fitting. SLIM serves as the foundation of the proposed BCA advisor, enabling configuration decisions across diverse LLM serving scenarios with limited profiling overhead. Beyond BCA, SLIM can also be applied to a broader range of optimization tasks, including autoscaling, system configuration exploration, and deployment optimization.

\subsection{Preliminaries}
\label{sec:model_tp}

We consider inference for decoder-only Transformer models characterized by the number of layers $L$, hidden dimension $D$, query heads $H_q$, key-value heads $H_{\mathrm{kv}}$, and head dimension $d_h$. Assuming $D=H_qd_h$, we define the total key/value projection dimension as
\begin{equation}
\label{eq:d_kv}
D_{\mathrm{kv}}
= H_{\mathrm{kv}}d_h
= D\frac{H_{\mathrm{kv}}}{H_q}.
\end{equation}

We estimate throughput and latency over a representative workload execution window. Specifically, the model considers a batch of size $B$ composed of requests with fixed input length $IL$ and output length $OL$. These parameters represent the average workload characteristics over the full execution and can be obtained from runtime profiling or external estimates. This abstraction simplifies the modeling of dynamic serving workloads while retaining predictive accuracy.

\subsection{Latency and Throughput}

As shown in Equation~\ref{eq:lat_model}, we model end-to-end (E2E) batch execution latency as the sum of the prefill time $T_{\mathrm{pre}}$ and the decode time $T_{\mathrm{dec}}$, which are described in the following sections, and a calibrated constant $t_{\mathrm{fixed}}$, which captures batch-level system overheads not explicitly modeled, such as request scheduling and runtime management. 
\begin{equation}
    E2E(IL,OL,B) = T_{\mathrm{pre}}+T_{\mathrm{dec}}+t_{\mathrm{fixed}}
    \label{eq:lat_model}
\end{equation}

Assuming that all requests in the batch complete simultaneously, this corresponds to their end-to-end latency. Total-token throughput is then computed by dividing the total number of processed tokens, $B(IL+OL)$, by the total execution time. 
\begin{equation}
    T(IL,OL,B)
    =
    \frac{B(IL+OL)}
    {E2E(IL,OL,B)}
    \label{eq:throughput_model}
\end{equation}

\subsection{Prefill Time}

We model the prefill time from its computational cost, as the prefill phase is predominantly compute-bound and memory-transfer overheads are comparatively small~\cite{agrawal2024taming}. As shown in Equation~\ref{eq:tprefill}, prefill time is estimated by dividing the total floating-point operations (FLOPs), $F_{\mathrm{pre}}$, by the accelerator's peak Tensor Core throughput, $F_{\max}$ (FLOP/s), scaled by a calibrated compute-efficiency factor, $\eta_c$. This factor captures the gap between theoretical peak performance and the effective throughput achieved in practice due to kernel-level and implementation-specific inefficiencies. 
\begin{equation}
    T_{\mathrm{pre}}
    =
    \frac{F_{\mathrm{pre}}}{\eta_c F_{\max}}
    \label{eq:tprefill}
\end{equation}

The total number of floating-point operations, $F_{\mathrm{pre}}$, is obtained by summing the self-attention and feed-forward network (FFN) operations across all Transformer blocks, counting each multiply-accumulate as two floating-point operations:
\begin{equation}
\begin{aligned}
F_{\mathrm{pre}}
= L \Big[
    &B*IL\left(
        4D^2 + 4DD_{\mathrm{kv}} + 4r_\mathrm{ff}D^2
    \right)
    + 4BIL^2D
\Big]
\end{aligned}
\label{eq:fprefill}
\end{equation}

The $4D^2$ term accounts for the query and output projections, $4DD_{\mathrm{kv}}$ for the key and value projections, and $4r_\mathrm{ff}D^2$ for the two FFN projections. Here, $r_{\mathrm{ff}}=D_{\mathrm{ff}}/D$ denotes the ratio between the FFN hidden dimension and the model dimension\textemdash for OPT models, $r_\mathrm{ff}=4$~\cite{vaswani2017attention}. Under the full-matrix attention formulation, computing $QK^\top$ and multiplying the resulting attention probabilities by $V$ each requires $2B\,\mathrm{IL}^2D$ FLOPs, yielding the $4B\,\mathrm{IL}^2D$ self-attention term. We use $D_{\mathrm{kv}}$ so that the projection cost reflects the reduced KV width of grouped-query attention (GQA) and multi-query attention (MQA) architectures~\cite{shazeer2019fast,ainslie2023gqa}.

\subsection{Decode Time}

For decode time, as shown in Equation~\ref{eq:tdecode}, we model two components: the dense decode time $T_{\mathrm{d}}$, and the KV-cache memory-transfer time, obtained by dividing the total memory traffic $M_{\mathrm{dec}}$ by the peak device bandwidth $BW_{max}$. Unlike prefill, decode becomes increasingly constrained by memory traffic under large-context workloads, as demonstrated in Section~\ref{sec:gpuprofiling}. We further introduce a calibrated decode-efficiency factor, $\eta_m$, and a small batch offset, $B_0$, to capture the nonlinear scaling of decode execution at small batch sizes.

\begin{equation}
\begin{aligned}
T_{\mathrm{dec}}
= \frac{B+B_0}{\eta_m}
\Bigg[T_{\mathrm{d}} + \frac{M_{\mathrm{dec}}}{BW_{\max}}
\Bigg]
\end{aligned}
\label{eq:tdecode}
\end{equation}

The dense-computation term $T_{\mathrm{d}}$ is modeled using an architecture-aware scaling formulation with calibrated coefficients $\alpha$ and $\gamma$:
\begin{equation}
\begin{aligned}
T_{\mathrm{d}} = \alpha OL + \gamma L\left(\frac{D}{1024}\right)^2 OL
\end{aligned}
\label{eq:cdecode}
\end{equation}

where the first term captures per-output-token overhead through the calibrated coefficient $\alpha$, accounting for residual costs such as decode-step scheduling and other lightweight operations. The second term models the execution time of the dense projection and feed-forward network (FFN) kernels through the calibrated coefficient $\gamma$. Its dependence on model architecture is expressed by the dominant scaling with the number of Transformer layers, $L$, and the square of the hidden dimension, $D$, which reflects the matrix-multiplication cost. Normalizing $D$ by 1024 improves numerical stability during parameter calibration.

Although both prefill and decode use calibrated parameters, their different parameterizations reflect their execution regimes. Prefill processes $BIL$ tokens concurrently, producing large matrix multiplications with sufficient parallelism for performance to be modeled as a calibrated fraction of $F_{\max}$. In contrast, decode processes only one new token per active request at each step. A fixed fraction of $F_{\max}$ is therefore less representative in this regime, so $\gamma$ directly absorbs the associated runtime overheads.

Finally, the memory-transfer workload, $M_{\mathrm{dec}}$, is modeled as
\begin{equation}
\begin{aligned}
M_{\mathrm{dec}} =
    2L D_{\mathrm{kv}} b_{\mathrm{kv}}
    C(IL,OL),
\end{aligned}
\label{eq:mdecode}
\end{equation}

which represents the total memory traffic generated by KV-cache reads during decoding. The parameter $D_{\mathrm{kv}}$ denotes the effective KV width, accounting for the reduced KV-cache size of GQA and MQA architectures. The function $C(IL,OL)$, defined in Equation~\ref{eq:context_scan}, captures the dependence of KV-cache accesses on sequence length: traffic scales linearly with the input length and quadratically with the output length because decoding repeatedly scans a growing KV cache. Since each context position stores both a key and a value vector, with each element occupying $b_{\mathrm{kv}}$ bytes, the total KV-cache traffic per request is given by $2L D_{\mathrm{kv}} b_{\mathrm{kv}} C(IL,OL)$. For standard multi-head attention (MHA),  $D_{\mathrm{kv}}=D$, whereas GQA and MQA reduce $D_{\mathrm{kv}}$ and therefore the corresponding memory traffic.
\begin{equation}
    C(IL,OL)
    = \sum_{j=1}^{OL} (IL+j)
    = OL(IL+1) + \frac{OL(OL-1)}{2}
    \label{eq:context_scan}
\end{equation}

\subsection{Evaluation Methodology}
\label{sec:eval_methodology}

\begin{figure*}[ht]
  \centering
  \includegraphics[width=1.0\linewidth]{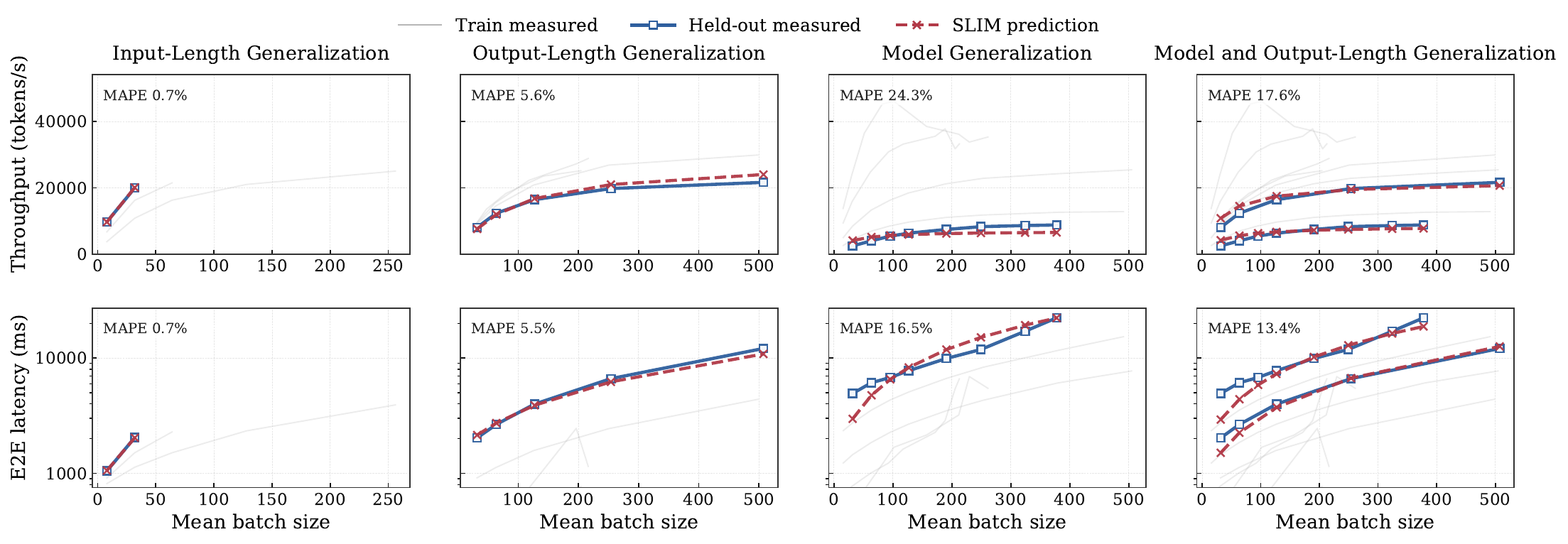}
\caption{Performance prediction accuracy when evaluating unseen input length, unseen output length, unseen model, and unseen model and output length. Each column reports one held-out scenario, with throughput shown on the top row and end-to-end latency on the bottom row as a function of mean batch size. Gray curves correspond to training measurements, while colored curves compare held-out measurements against SLIM predictions.}
\label{fig:chart_mean_batch_size_models_outputlength_heldout_SLIM_with_estimation}
\end{figure*}

We evaluate SLIM using measured vLLM serving traces in two settings: accuracy on held-out batch sizes from the fitted serving curves, and generalization to unseen model sizes, input lengths, and output lengths without refitting. We estimate the parameters of the semi-analytical model on the training set by minimizing the sum of squared relative errors between measured and predicted throughput using nonlinear least squares. Each observation corresponds to a serving configuration defined by its model, input length, output length, and batch size. We evaluate prediction accuracy by reporting the mean absolute percentage error (MAPE) between the profiled and estimated total-token throughput and E2E latency.

In our evaluation, we consider five generalization scenarios, summarized in Table~\ref{tab:heldout_validation_settings}. First, the same-curve evaluation holds out batch-size points from each measured throughput curve. For every model and output-length curve, we use three points in the pre-knee regime for fitting and the remaining points for validation. This tests whether the estimator can recover the throughput-saturation shape from sparse profiling points on the same curve. Second, the model-transfer evaluation fixes the output length at OL=384 and fits on smaller OPT models, using OPT-125M, OPT-350M, OPT-1.3B, and OPT-2.7B for training and OPT-6.7B for validation. Third, the output-length-transfer evaluation fixes the input length and fits on shorter generations, using OL $\in {32,128,256}$ for training and OL=512 for validation. Fourth, the input-length-transfer evaluation uses OPT-1.3B as the fixed model and the output length to OL=256, fits on IL $\in {128,512}$, and validates on IL=1024. Finally, the combined setting fits a single global parameterization on the union of the model-transfer and output-length-transfer training sets and evaluates this single fitted model on OPT-6.7B at OL=512, a configuration that is simultaneously unseen in both model scale and output length. This evaluation scenario tests whether one fitted model can generalize across both model scale and generation length.

We compare SLIM against two higher-level predictors adapted to our profiling setting. First, we construct LLMVisor-Agg, an adaptation of LLMVisor \cite{jinllmvisor}, which uses piecewise fitted latency functions for prefill and decode, with aggregate token-count features for batch size, context tokens, and squared prefill length. The original LLMVisor trains its piecewise functions from per-engine-step execution times. Our profiling interface instead provides one mean TTFT and one mean ITL measurement for each $(\mathrm{model}, B, IL, OL)$ configuration.  We therefore fit the prefill and decode functions to mean TTFT and mean ITL, respectively, and reconstruct end-to-end latency. We additionally adapt the out-of-domain linear-regression approach of Imai et al. \cite{imai2024predicting}, labeled as IMAI OOD-LR. This LR baseline predicts end-to-end latency from workload variables and numerical model characteristics, including batch size, input and output lengths, hidden dimension, layer and attention-head counts, and interaction features. Throughput is then derived from the predicted latency and total token count.

\begin{table}[t]
\centering
\caption{Held-out validation settings used for SLIM's prediction-accuracy
evaluation. $B$ is the defined batch size while $IL$/$OL$ denote input and output length, respectively.}
\label{tab:heldout_validation_settings}
\footnotesize
\setlength{\tabcolsep}{3pt}
\renewcommand{\arraystretch}{1.15}
\begin{tabular}{@{}p{0.16\columnwidth}p{0.40\columnwidth}p{0.34\columnwidth}@{}}
\toprule
\textbf{Setting} & \textbf{Training set} & \textbf{Validation target} \\
\midrule
Same curve &
All fitted model--$OL$ curves, $B\in\{1,4,8\}$ &
Same curves, $B\in\{2,16,32,64,128,512\}$ \\
Model &
OPT-125M, OPT-350M, OPT-1.3B, OPT-2.7B at $OL=384$ &
OPT-6.7B, $OL=384$ \\
Input length &
OPT-1.3B, $IL\in\{128,512\}$ &
OPT-1.3B, $IL=1024$ \\
Output length &
OPT-1.3B, $OL\in\{32,128,256\}$ &
OPT-1.3B, $OL=512$ \\
Model + output length &
Union of the \emph{Model} and \emph{Output length} training sets above &
OPT-6.7B, $OL=512$ \\
\bottomrule
\end{tabular}
\end{table}

\subsection{Experimental Evaluation}
\label{sec:evaluation}

\subsubsection{Model Prediction Accuracy}
\label{sec:model_accuracy}

Figure~\ref{fig:chart_mean_batch_size_models_outputlength_heldout_SLIM_with_estimation} compares measured and predicted throughput-latency curves under held-out serving configurations. Despite being calibrated from sparse measurements on smaller models and shorter generation lengths, SLIM captures the saturation trend for unseen input lengths, output lengths and model scales. Input- and output-length generalization scenarios achieve throughput MAPE values below 6\%, because these scenarios primarily extend the same computation and KV-cache scan mechanisms represented by the model. Model generalization is more challenging, yielding a throughput MAPE of 24.3\%, because changing the model simultaneously affects architectural characteristics, kernel behavior, and runtime overheads that are not fully captured analytically. Nevertheless, even in this setting, SLIM reconstructs the overall throughput-saturation curve, supporting its use to reduce the profiling effort required to identify efficient serving configurations.

\begin{figure}[ht]
  \centering
  \vspace{-5pt}
  \includegraphics[width=1.0\linewidth]{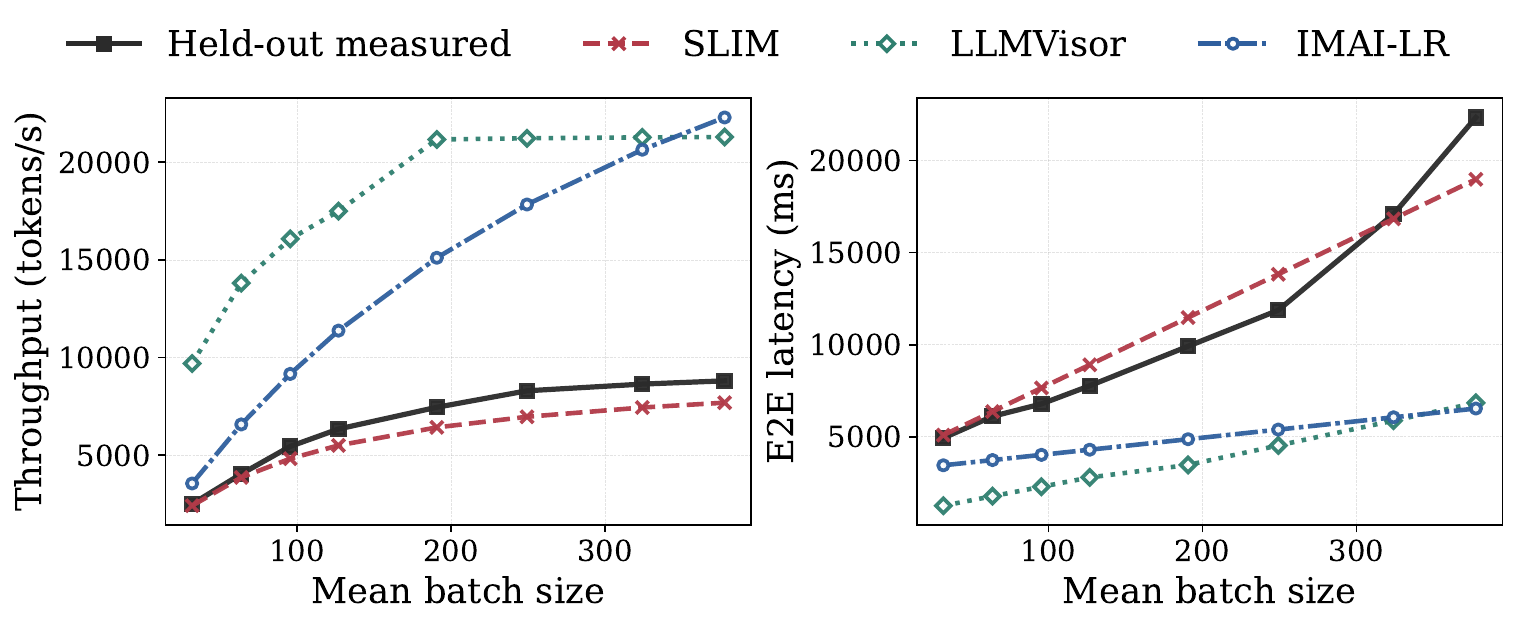}
\caption{Held-out prediction for an unseen model. The estimators are fitted on OPT-125M, OPT-350M, OPT-1.3B, and OPT-2.7B at OL=384, then evaluated on OPT-6.7B at OL=384.}
  \label{fig:comparison_fit_models}
\end{figure}

Table~\ref{tab:modeling_heldout_results_simple_tp} quantifies the resulting prediction accuracy and compares SLIM with the two baselines. LLMVisor captures architectural and hardware effects only indirectly through fitted coefficients, but performs well when extrapolating across output lengths. Imai et al. encode model characteristics as regression features and achieve higher accuracy under input-length generalization. In contrast, SLIM explicitly models the decode-time KV-cache traffic responsible for throughput saturation and the model-dependent dense-computation cost, yielding more accurate predictions under model and output length generalization. Across all scenarios, SLIM reduces average MAPE by 79.3\% relative to the two baselines. This advantage is preserved in the most challenging scenario, where both the model and output length are unseen during training, with an average MAPE reduction of 78.7\% over the baselines. Figure~\ref{fig:comparison_fit_models} further illustrates this trend, where baseline models generalize poorly to an unseen LLM, whereas SLIM closely follows the profiled throughput and latency curves.

\begin{table*}[t]
\centering
\caption{Prediction error under held-out batch-size, sequence-length, and model-scale configurations. We report MAPE (\%) for total throughput and end-to-end latency; lower values indicate higher prediction accuracy.}
\label{tab:modeling_heldout_results_simple_tp}
\footnotesize
\begin{tabularx}{\textwidth}{>{\raggedright\arraybackslash}X
                            >{\raggedright\arraybackslash}X
                            c c c c c c}
\toprule
\multirow{2}{*}{Held-out setting}
& \multirow{2}{*}{Validation target}
& \multicolumn{2}{c}{SLIM (ours)} & \multicolumn{2}{c}{LLMVisor-Agg} & \multicolumn{2}{c}{IMAI OOD-LR} \\
\cmidrule(lr){3-4}
\cmidrule(lr){5-6}
\cmidrule(lr){7-8}
&
& Throughput & E2E latency & Throughput & E2E latency & Throughput & E2E latency \\
\midrule
Same curve  & Held-out batch sizes & \textbf{15.1} & \textbf{17.5} & 55.8 & 81.2 & 55.3 & 42.9 \\

Input length  & IL=1024, OL=256 & \textbf{0.7} & \textbf{0.7} & 68.1 & 255.2 & 1.2 & 1.2 \\

Model  & OPT-6.7B, OL=384 & \textbf{24.3} & \textbf{16.5} & 141.8 & 59.9 & 39.9 & 72.8 \\

Output length  & OPT-1.3B, OL=512 & \textbf{5.6} & \textbf{5.5} & 7.7 & 8.9 & 38.9 & 25.9 \\

Model + output length  & Combined & \textbf{17.6} & \textbf{13.4} & 95.8 & 43.7 & 91.7 & 60.2 \\
\bottomrule
\end{tabularx}

\end{table*}

\subsubsection{SLIM Ablation}
\label{sec:model_components}

\begin{table}[t]
\centering
\footnotesize
\caption{Component ablation of SLIM under the held-out model-transfer and output-length-transfer settings. Values are MAPE (\%) relative to the measured throughput and end-to-end latency curves; lower is better.}
\label{tab:SLIM_ablation}
\begin{tabular}{@{}llrr@{}}
\toprule
Transfer setting & Variant & \multicolumn{2}{c}{MAPE (\%)} \\
\cmidrule(lr){3-4}
& & Throughput & E2E latency \\
\midrule
Model & Full SLIM & 24.3 & \textbf{16.5} \\
 & w/o KV scan & \textbf{23.5} & 20.9 \\
 & w/o dense decode & 31.6 & 18.9 \\
\addlinespace[2pt]
Output length & Full SLIM & \textbf{5.6} & \textbf{5.5} \\
 & w/o KV scan & 33.0 & 24.8 \\
 & w/o dense decode & 8.1 & 8.2 \\
\bottomrule
\end{tabular}
\end{table}

We assess the contribution of SLIM's two decode components through the ablation study summarized in Table~\ref{tab:SLIM_ablation}. Removing the memory-transfer term, $M_{\mathrm{dec}}$, substantially degrades output-length generalization, increasing throughput MAPE from 5.6\% to 33.0\%. In contrast, removing the dense computation term, $T_{\mathrm{d}}$, primarily affects model generalization, increasing the throughput MAPE from 24.3\% to 31.6\%. These results show that $M_{\mathrm{dec}}$ effectively captures the context-dependent memory traffic incurred during decoding, whereas $T_{\mathrm{d}}$ captures the architecture-dependent dense-computation cost. The two terms therefore make complementary contributions to SLIM’s predictive accuracy.

\subsubsection{Large Model Generalization and Limitations}
\label{sec:scale_parallelism}

\begin{figure}[ht]
  \centering
  \vspace{-5pt}
  \includegraphics[width=1.0\linewidth]{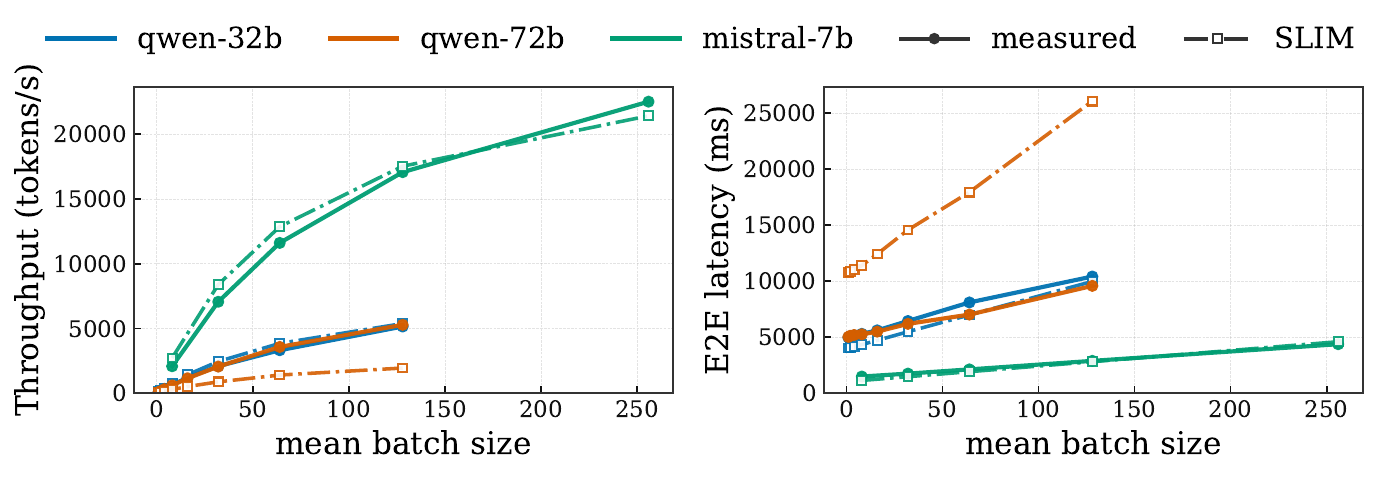}
  \caption{Predicted and measured throughput and end-to-end latency for large
  Qwen models under tensor-parallel serving and for Mistral-7B as an unseen grouped-query-attention architecture. Solid curves report measured performance, while dash-dotted curves report SLIM predictions using coefficients fitted only on single-GPU OPT measurements.}
  \label{fig:qwen_tp_prediction}
\end{figure}

We further evaluate SLIM’s ability to generalize to larger unseen models and tensor-parallel deployments. Figure~\ref{fig:qwen_tp_prediction} compares measured throughput and end-to-end latency with SLIM predictions for Qwen-32B and Qwen-72B, served on two and four H100 GPUs, respectively. The results show that SLIM captures the throughput-saturation trend as batch size increases, but overestimates the performance gap between the two models. Under tensor parallelism, Qwen-72B is sharded across four GPUs, resulting in a per-GPU parameter footprint of approximately 16.5~GB, close to the roughly 15~GB per GPU of Qwen-32B. Their measured throughput curves are therefore similar, suggesting that per-GPU workload dominates and that tensor-parallel communication overhead is limited in this setting. Because SLIM does not explicitly model tensor-parallel sharding, it treats Qwen-72B as a uniformly larger model and consequently predicts a larger throughput reduction than observed.

We also include in Figure~\ref{fig:qwen_tp_prediction} the Mistral-7B model, whose architecture differs from the OPT models used to calibrate SLIM's coefficients. In particular, Mistral-7B employs grouped-query attention rather than standard multi-head attention. By accounting for its reduced KV width through $D_{\mathrm{kv}}$ in Equation \ref{eq:d_kv}, SLIM closely follows the measured throughput and latency curves across the evaluated batch sizes, including their characteristic saturating behavior. These results suggest that SLIM generalizes not only to unseen model scales within the OPT family but also to an unseen attention architecture.

\section{Model-Guided Batching Configuration}
\label{sec:advisor}

In this section, we introduce the \textbf{SLIM-guided Batching Configuration Advisor (SLIM-BCA)}, an offline, model-driven advisor that recommends a batch size cap $B_\mathrm{cap}$ by jointly accounting for the predicted throughput knee and a latency SLO. Rather than exhaustively profiling the serving system, SLIM-BCA evaluates candidate configurations using SLIM’s fitted throughput and latency estimates. As shown in Section~\ref{sec:gpuprofiling}, increasing the batch size beyond a deployment-specific knee provides diminishing throughput returns once decode-time attention approaches DRAM-bandwidth saturation, while latency and KV-cache requirements continue to grow. SLIM-BCA therefore selects the largest admissible batch size that remains within the efficient pre-saturation region and satisfies the SLO. This operating point preserves near-maximum throughput while avoiding unnecessary KV-cache allocation, thereby releasing memory capacity that can be used to serve additional requests, model instances, or co-located workloads.

\subsection{Formulation}

Formally, Equation~\ref{eq:Bopt} defines $B_{\mathrm{cap}}$ as the batch size that maximizes throughput among configurations satisfying the latency SLO and minimum batching-efficiency constraint. SLIM-BCA uses the performance model to estimate $T(IL,OL,B)$ directly from model, workload, and accelerator characteristics. Then, given a workload-dependent $IL$ and $OL$, the advisor searches over batch size candidates $B$ and returns the upper-bound estimate $B_{\mathrm{cap}}$ for the specified SLO and $\epsilon$.

\begin{equation}\label{eq:Bopt}
\begin{aligned}
B_{\mathrm{cap}}(IL,OL)
&= \arg\max_{B}\; T(IL,OL,B) \\
\text{subject to} \quad
& \left\{
\begin{array}{l}
E2E(IL,OL,B) \le \textsc{SLO}, \\[3pt]
\dfrac{T(IL,OL,B)}{B \cdot T(IL,OL,1)} > \epsilon .
\end{array}
\right.
\end{aligned}
\end{equation}

The search in Equation~\ref{eq:Bopt} is subject to two constraints. First, the predicted latency $E2E(IL,OL,B)$ must not exceed the specified SLO. Second, the ratio $T(IL,OL,B)/(B \cdot T(IL,OL,1))$, which is the throughput at batch size $B$ relative to what would be achieved if performance scaled linearly with $B=1$, must remain above a user-defined threshold $\epsilon$. This prevents the advisor from selecting a batch size that is in the plateau regime. Both SLO and $\epsilon$ are user-defined parameters that trade off latency guarantees against batching efficiency. In our experiments, we define a tight SLO as $2*E2E(2,128,32)$, and a loose SLO as $4*E2E(2,128,32)$, which is four times the latency at a specific profiling point.

\subsection{Evaluation of SLIM-BCA}
We evaluate SLIM-BCA along three dimensions: the quality of its recommendations relative to a fully profiled ground truth (Section~\ref{sec:bca-performance}), the reduction in profiling effort required to obtain those recommendations (Section~\ref{sec:bca-cost}), and the resulting savings in KV-cache allocation (Section~\ref{sec:bca-memory}).

\subsubsection{Profiling Performance}
\label{sec:bca-performance}

\begin{table}[t]
\centering
\caption{SLIM-BCA decision variation as more profiling curves are added. Throughput variation and SLO violation are evaluated against the full-profile BCA decision on measured data and $\epsilon$=0.1.}
\label{tab:aca_quality}
\footnotesize
\begin{tabularx}{\columnwidth}{
    >{\raggedright\arraybackslash}p{0.18\columnwidth}
    >{\raggedright\arraybackslash}p{0.28\columnwidth}
    c c c}
\toprule
Target & Train & N & Thr. Var & SLO Viol. \\
\midrule
OL-512 & 16+32 & 5 & 34.6\% & 0.0\% \\
OL-512 & 16+32+48 & 8 & 0.0\% & 0.0\% \\
OL-512 & 16+32+48+64 & 13 & 0.0\% & 0.0\% \\
OPT-6.7B & 125M & 9 & -28.4\% & 6.9\% \\
OPT-6.7B & 125M+350M & 18 & 15.1\% & 0.0\% \\
OPT-6.7B & 125M+350M+1.3B & 29 & 15.1\% & 0.0\% \\
\bottomrule
\end{tabularx}
\end{table}

Table~\ref{tab:aca_quality} evaluates SLIM-BCA as a low-cost alternative to full offline profiling, which requires profiling the system for each candidate batch size. Instead, SLIM-BCA fits the performance model on a small set of profiling points and uses the predicted curve to find the optimal batch size with the same BCA decision rule. We then compare the resulting batch size recommendation against full-profile BCA with $\epsilon=0.1$. Following the same experimental set-up as Section \ref{sec:modeling}, we evaluate performance in scenarios with unseen models and output lengths while calibrating SLIM only on the smallest models and shortest output lengths, which incur the lowest profiling cost.

For output-length transfer, profiling only $OL$ 16 and 32 leads to a 34.6\% throughput variation on OL-512 when comparing the full-profiling batch size decision, whereas adding OL-48 removes this discrepancy. For model transfer to OPT-6.7B, profiling on OPT-125M alone results in a 28.4\% throughput discrepancy relative to the full-profile decision and a 6.9\% SLO violation. Adding OPT-350M and OPT-1.3B reduces the discrepancy to 15.1\% and removes the SLO violation entirely. Overall, these results show that a small number of profiled curves is sufficient for SLIM-BCA to closely recover the operating point selected by full-profile BCA. Because the predicted curves preserve the shape of the fully profiled curves, SLIM-BCA produces similar batch-size recommendations. Nevertheless, the SLO violation observed under model generalization evaluation highlights the residual risk of relying on predicted latency rather than directly measuring the target configuration.

\begin{figure}[ht] 
\centering 
  \includegraphics[width=1.0\linewidth]{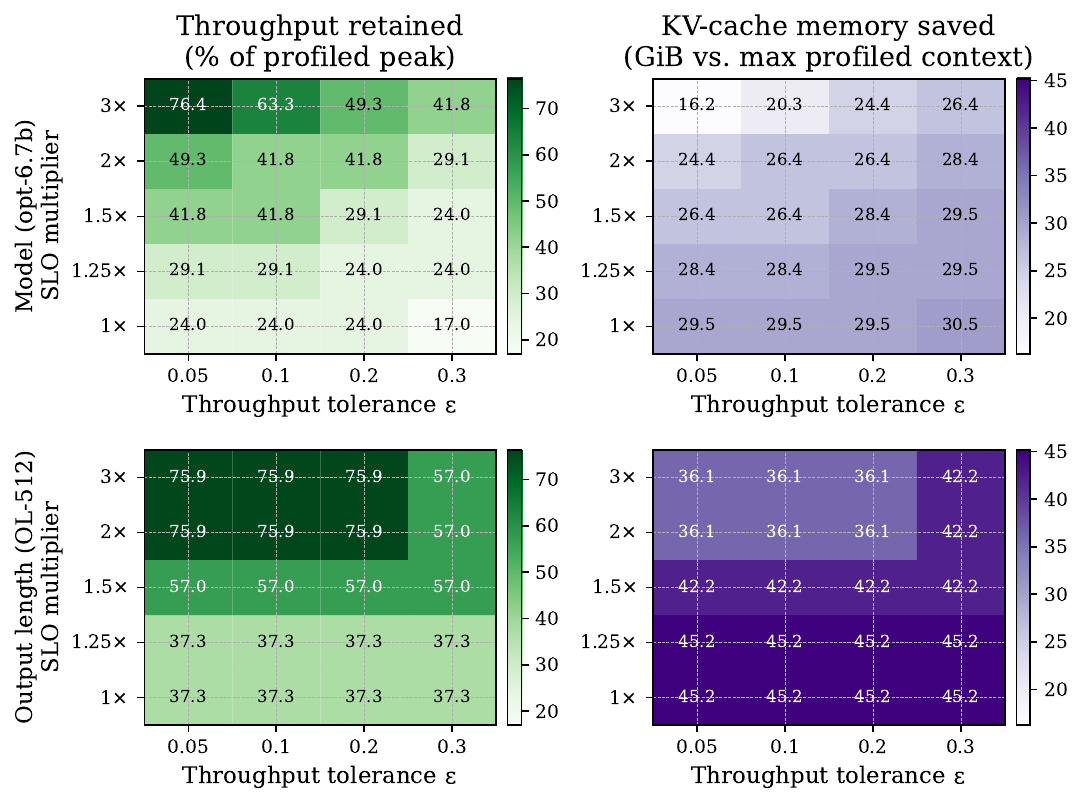}
\caption{Sensitivity of SLIM-BCA to the throughput-tolerance threshold $\epsilon$ and latency-SLO multiplier under model generalization (top) and output-length generalization (bottom). The left column reports throughput retained relative to the profiled peak, while the right column reports KV-cache memory saved relative to the maximum profiled context.}
\label{fig:sensitivity}
\end{figure}

Figure~\ref{fig:sensitivity} shows the trade-off induced by SLIM-BCA’s throughput threshold and latency constraint. Under model generalization, looser SLOs allow higher retained throughput, whereas increasing $\epsilon$ selects more conservative batch-sizes and yields larger memory savings. Under output-length generalization, the recommendation remains unchanged for $\epsilon\leq0.2$ across most SLO levels, indicating limited sensitivity to the threshold. Only $\epsilon=0.3$ reduces throughput under the loosest SLOs, confirming that SLIM-BCA is generally robust while exposing the expected throughput–memory trade-off.

\subsubsection{Profiling-cost Savings}
\label{sec:bca-cost}

Table \ref{tab:aca_efficiency} quantifies the profiling effort avoided by fitting SLIM instead of exhaustively profiling every serving configuration. For each held-out setting, \emph{Prof.} reports the number of configurations actually profiled to fit SLIM, whereas \emph{Pred.} reports the number of held-out configurations whose throughput and latency are obtained from SLIM's predictions. \emph{Pt. save} denotes the resulting fraction of configurations that do not need to be profiled directly. Profiling-point savings range from 24.2\% in the model-transfer setting, where 8 of 33 evaluated configurations are predicted by SLIM rather than directly profiled, to 66.7\% in the same-curve setting, where only 3 of the 9 batch-size points on each curve need to be measured.

Reducing the number of profiled configurations also reduces total profiling time. Because we evaluate SLIM-BCA under generalization to larger models and longer output lengths, the held-out configurations are often the most expensive to profile. For instance, by profiling 25.0\% fewer configurations we reduce profiling time by 59.9\%, because the held-out configuration (OL\,=\,512) is also the most expensive, as longer generations require proportionally more decode steps. This indicates that SLIM's benefit extends beyond the reduction in the number of profiling runs, as it specifically avoids the most expensive ones, which are concentrated in larger models and longer generations.

\begin{table}[t]
\centering
\caption{BCA profiling cost and recommendation quality when fitting SLIM and predicting held-out throughput curves instead of fully profiling every curve.}
\label{tab:aca_efficiency}
\footnotesize
\begin{tabular}{>{\raggedright\arraybackslash}p{0.32\columnwidth}rrrr}
\toprule
Setting & Prof. & Pred. & Pt. save & Time save \\
\midrule
Same curve & 3 & 6 & 66.7\% & 56.9\% \\
O. length & 15 & 5 & 25.0\% & 59.9\% \\
Model & 25 & 8 & 24.2\% & 27.2\% \\
Model + O. length & 40 & 13 & 24.5\% & 31.4\% \\
\bottomrule
\end{tabular}
\end{table}

\subsubsection{Memory Savings}
\label{sec:bca-memory}

\begin{figure}[ht]
  \centering
  \includegraphics[width=0.9\linewidth]{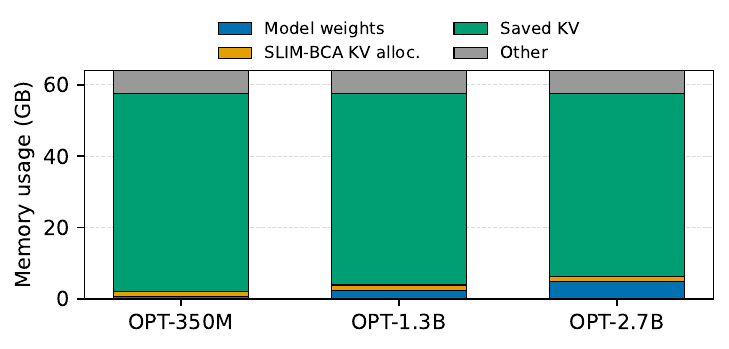}
  \caption{Memory usage distribution for each model size in our 64GB GPU environment, considering $B_{cap}$ under a strict SLO and $\epsilon = 0.1$. By default, vLLM allocates 90\% of available memory, leaving 10\% for the other processes (Other).}
  \label{fig:gpu_memory_distribution_with_bopt_real_data}
\end{figure}

Common LLM serving frameworks such as vLLM reserve the maximum available GPU memory for serving by default. While this is often a good practice for larger models, we show in Figure~\ref{fig:gpu_memory_distribution_with_bopt_real_data} the memory savings achieved by SLIM-BCA when serving smaller models on a high-end GPU. Instead of reserving the full KV-cache budget, SLIM-BCA allocates only enough KV-cache memory to support the batch size proposed by the model. In our setup, this corresponds to 1.52~GB for OPT-350M, 6.09~GB for OPT-1.3B, and 5.08~GB for OPT-2.7B, leaving 55.42~GB, 49.08~GB, and 47.49~GB of the default KV-cache allocation unused, respectively. These results show that SLIM-BCA can identify substantial memory savings while preserving its selected operating point. 

\section{Discussion}
\label{sec:discussion}

In this work, we identify the GPU bottlenecks underlying throughput saturation in LLM inference with large active contexts. Decode-time attention maintains nearly constant arithmetic intensity as batch size or context length increases, causing KV-cache traffic to approach the DRAM-bandwidth limit while compute resources remain underutilized. This is consistent with the widening gap between accelerator compute and memory bandwidth, where peak compute has historically scaled much faster than DRAM and interconnect bandwidth~\cite{gholami2024ai}. We therefore expect attention-bandwidth bottlenecks to persist across future accelerators unless mitigated by substantially higher DRAM/HBM bandwidth or reduced KV-cache traffic. The low cache hit rates and increasing long-scoreboard stalls further indicate that the bottleneck arises from streaming KV-cache data with limited temporal reuse, rather than from an isolated cache-management inefficiency. Our kernel-level profiling is limited to Mistral-7B and Granite-8B on a single H100 GPU, so validation across accelerators and KV-cache formats remains future work. Nevertheless, the same mechanism is expected to occur across conventional decoder-only attention architectures, as the attention kernel’s computation-to-memory ratio is largely determined by the model’s attention structure itself.

We propose SLIM as a lightweight semi-analytical model that translates these hardware observations into throughput and latency estimates. Its explicit memory-transfer term supports accurate generalization across input and output lengths, while a fitted dense-decode correction captures model-scale effects. Consequently, SLIM achieves a MAPE below 6\% when generalizing to unseen sequence lengths by extrapolating the same memory behavior, while generalization across models is more challenging, yielding a MAPE of 24.3\% due to differences in model architecture and runtime overheads. SLIM aims to minimize profiling rather than eliminate it entirely, since costs such as kernel launches, framework scheduling, and non-attention operators are difficult to model from first principles. The current formulation targets decoder-only autoregressive inference, reducing MAPE by more than 70\% relative to the evaluated baselines, but would require recalibration to account for recent optimizations such as quantized KV caches, speculative decoding, and prefix caching. In most cases, however, these changes affect existing terms rather than the overall structure, as quantization reduces $b_{kv}$, prefix caching reduces the effective attended input context, and speculative decoding changes the effective per-step workload. 

Finally, we integrate SLIM with our proposed Batching Configuration Advisor (BCA) to select efficient batch-size and KV-cache configurations under latency constraints. Although BCA currently operates as an offline configuration advisor rather than an online scheduler and requires estimates of incoming request lengths, using average workload characteristics\textemdash readily obtainable in production environments\textemdash proves effective, reducing GPU memory usage by up to 55 GB for the evaluated OPT models with only minor throughput degradation. This makes it suitable for capacity planning, complementing reactive schedulers and autoscalers that can track its recommended operating point. In practice, BCA recommendations should include conservative throughput and latency margins, or be combined with lightweight runtime monitoring that adjusts the selected configuration when observed latency deviates from predictions. The reported memory reductions should similarly be interpreted as potential KV-cache reservation savings rather than automatically reclaimed capacity. Finally, because SLIM is calibrated per accelerator, applying BCA across heterogeneous GPUs requires re-calibration rather than direct reuse. Integrating these offline recommendations with adaptive runtime control is a natural next step toward multi-model orchestration under changing workloads and service-level objectives.

\section{Conclusion}\label{sec:conclusion}

In this work, we conduct an in-depth GPU analysis to identify the performance bottlenecks responsible for throughput saturation in large active-context LLM inference. Our findings show that decode-time attention remains memory-bound, with KV-cache traffic driving DRAM-bandwidth saturation while substantial compute capacity remains underutilized. Building on this characterization, we introduce SLIM, a lightweight semi-analytical model that predicts throughput and latency across different model scales, batch sizes, and sequence lengths using only sparse profiling. We further instantiate SLIM through BCA, which selects efficient batch and KV-cache configurations under latency constraints, reducing unnecessary memory allocation while preserving near-maximum throughput among SLO-feasible configurations. Overall, our results provide a hardware-grounded explanation of throughput saturation and a practical methodology for configuring LLM-serving deployments to improve GPU resource efficiency.

\section*{Acknowledgments}

This work has been partially financed by grant agreement EU-HORIZON GA.101296742 (OptimAIse). Also, it has been partially financed by Generalitat de Catalunya (AGAUR) under grant agreement 2021-SGR-00478, by Severo Ochoa Center of Excellence CEX-2021-001148-S-20-3, and by the Spanish Ministry of Science (MICINN), the Research State Agency (AEI) and European Regional Development Funds (ERDF/FEDER) under grant agreement PID2021-126248OB-I00, MCIN/AEI/10.13039/ 501100011033/ FEDER, UE.

The authors used AI generative models to assist with text rephrasing and stylistic refinement. All content was originally written and subsequently reviewed and validated by the authors.

\bibliographystyle{elsarticle-harv}
\bibliography{ms}


\end{document}